\documentclass[11pt]{article}
\pdfoutput=1
\usepackage{hyperref}
\begin{document}
\title{Ants as Fluids: Physics-Inspired Biology}
\author{Micah Streiff, Nathan Mlot, Sho Shinotsuka, Alex Alexeev, David Hu\\
\\\vspace{6pt} George W. Woodruff School of Mechanical Engineering, \\ Georgia Institute of Technology, Atlanta, GA 30332, USA}
\maketitle
%% The abstract (in this file, and that submitted as text to arXiv) should include the exact phrase
%% "fluid dynamics video" or "fluid dynamics videos"
\begin{abstract}
Fire ants use their claws to grip diverse surfaces, including each other.  As a result of their mutual adhesion and large numbers, ant colonies flow like inanimate fluids.  In this sequence of films, we demonstrate how ants behave similarly to the spreading of drops, the capillary rise of menisci, and gravity-driven flow down a wall.  By emulating the flow of fluids, ant colonies can remain united under stressful conditions. 
\end{abstract}
% main text
%\section{Introduction}

%
\end{document}